\pdfoutput=1
\RequirePackage{ifpdf}
\ifpdf 
\documentclass[pdftex]{sigma}
\else
\documentclass{sigma}
\fi

\numberwithin{equation}{section}

\begin{document}

\allowdisplaybreaks

\renewcommand{\thefootnote}{$\star$}

\renewcommand{\PaperNumber}{051}

\FirstPageHeading

\ShortArticleName{A Two-Component Generalization of the Integrable rdDym Equation}

\ArticleName{A Two-Component Generalization\\ of the Integrable rdDym Equation\footnote{This
paper is a contribution to the Special Issue ``Geometrical Methods in Mathematical Physics''. The full collection is available at \href{http://www.emis.de/journals/SIGMA/GMMP2012.html}{http://www.emis.de/journals/SIGMA/GMMP2012.html}}}

\Author{Oleg I. MOROZOV}

\AuthorNameForHeading{O.I.~Morozov}

\Address{Institute of Mathematics and Statistics, University of Troms\o, Troms\o~90-37, Norway}
\Email{\href{mailto:Oleg.Morozov@uit.no}{Oleg.Morozov@uit.no}}

\ArticleDates{Received May 26, 2012, in f\/inal form August 09, 2012; Published online August 11, 2012}

\Abstract{We f\/ind a two-component generalization of the integrable case of rdDym equation. The reductions of this system
include the general rdDym equation, the Boyer--Finley equation, and the deformed Boyer--Finley equation.
Also we f\/ind a B\"acklund transformation between our generalization and Bodganov's two-component
generalization of the universal hierarchy equation.}


\Classification{35A30; 58H05; 58J70}

\renewcommand{\thefootnote}{\arabic{footnote}}
\setcounter{footnote}{0}

\section{Introduction}

Recent papers  \cite{Bogdanov1, Dunajski2002, ManakovSantini2006}
provide two-component generalizations for
the hyper-CR Einstein--Weil structure equation \cite{Dunajski2004, Pavlov2003}
\begin{gather}
s_{yy} = s_{tx}+s_y s_{xx}-s_x s_{xy},
\label{Pavlov_eq}
\end{gather}
Pleba\'{n}ski's second heavenly equation \cite{Plebanski1975}
\begin{gather}
s_{xz} = s_{ty}+s_{xx} s_{yy}-s_{xy}^2
\label{Plebanski_II_eq}
\end{gather}
and the universal hierarchy equation \cite{MartinezAlonsoShabat2002,MartinezAlonsoShabat2004, Pavlov2003}
\begin{gather}
s_{xx} = s_x  s_{ty}-s_t s_{xy}.
\label{universal_hierarchy_eq}
\end{gather}
Namely, equations (\ref{Pavlov_eq})--(\ref{universal_hierarchy_eq}) appear from systems
\begin{gather}
\begin{split}
& s_{yy}  = s_{tx}+(s_y+r) s_{xx}-s_x s_{xy},
\\
& r_{yy}  = r_{tx}+(s_y+r) r_{xx}-s_x r_{xy} + r_x^2;
\end{split}
\label{Manakov_Santini_system}
\\
\begin{split}
& s_{xz}  =  s_{ty}+s_{xx} s_{yy}-s_{xy}^2+r,
\\
& r_{xz}  =  r_{ty}+ s_{yy} r_{xx}+s_{xx} r_{yy}-2 s_{xy} r_{xy},
\end{split}
\label{Dunajski_system}
\end{gather}
and
\begin{gather}
\begin{split}
s_{xx}  =  e^r (s_x s_{ty}-s_t s_{xy}),
\\
\left(e^{-r}\right)_{xx}  =  s_x r_{ty}-s_t r_{xy},
\end{split}
\label{Bogdanov_system}
\end{gather}
respectively, by substituting for $r=0$.
Other reductions for (\ref{Manakov_Santini_system}) are found in \cite{Dunajski2008,ManakovSantini2006}:
when $u=0$, system (\ref{Manakov_Santini_system}) gives the Khokhlov--Zabolotskaya (or dispersionless
Kadomtsev--Petviashvili) equation
\begin{gather*}
v_{yy} = v_{tx}+v v_{xx}+ v_x^2,
\end{gather*}
while substituting for $v=u_x$ in  (\ref{Manakov_Santini_system}) produces the normal form
\begin{gather*}
u_{yy} = u_{tx}+(u_x+u_y) u_{xx} - u_x u_{xy},
\end{gather*}
for the family of equations studied in \cite{Dunajski2008}. Also, we note the reduction $v=u_y$ for~system (\ref{Manakov_Santini_system}). This reduction yields equation
\begin{gather*}
u_{yy} = u_{tx}-u_x u_{xy}
\end{gather*}
studied in \cite{FerapontovMoroSokolov2008, Krichever1994,ManasMedinaMartinez2006,Pavlov2004}.

As it was shown in~\cite{Bogdanov1}, the reduction $s=x$ for system (\ref{Bogdanov_system}) gives the Boyer--Finley equation
\begin{gather}
r_{ty} = \left(e^{-r}\right)_{xx}.
\label{Boyer_Finley_eq}
\end{gather}

The purpose of  the present paper is to introduce the two-component generalization for equation
\begin{gather}
u_{ty} = u_x u_{xy}-u_y u_{xx},
\label{special_rdDym_eq}
\end{gather}
which is integrable in the following sense: it has the dif\/ferential covering
\cite{VK1999,KrasilshchikLychaginVinogradov1986,KrasilshchikVinogradov1984,
KrasilshchikVinogradov1989}
\begin{gather}
p_t  = (u_x-\lambda) p_x,
\qquad
p_y  =  \lambda^{-1} u_y p_x
\label{M_1_covering}
\end{gather}
containing the non-removable parameter $\lambda\not = 0$~\cite{Morozov2009}.
We show that reductions of the generalization
include the general  $r$-th dispersionless Dym equation~\cite{Blaszak2002}
\begin{gather}
u_{ty} = u_x  u_{xy}+\kappa u_y u_{xx},
\label{general_rdDym_eq}
\end{gather}
the Boyer--Finley equation (\ref{Boyer_Finley_eq}), and the deformed Boyer--Finley equation.
Also we f\/ind a~B\"ack\-lund transformation between our generalization and Bodganov's two-component
gene\-ra\-li\-za\-tion~(\ref{Bogdanov_system})
of the universal hierarchy equation~(\ref{universal_hierarchy_eq}).

\section{The two-component generalization}

Along with the covering (\ref{M_1_covering}) equation (\ref{special_rdDym_eq}) has the covering
\begin{gather}
q_t  = (u_x-q) q_x,
\qquad
q_y  =  u_y q^{-1} q_x,
\label{M_2_covering}
\end{gather}
which can be obtained by the method of \cite{Morozov2009}.
While the coverings (\ref{M_1_covering}) and (\ref{M_2_covering}) are not
equivalent w.r.t.\ the pseudo-group of contact transformations, (\ref{M_2_covering}) can be
derived from~(\ref{M_1_covering}) by the following procedure, see, e.g., \cite{PavlovChangChen2009}.
We consider the function $p=p(t,x,y)$ from (\ref{M_1_covering}) to be def\/ined implicitly by the
equation $q(t,x,y,p(t,x,y)) = \lambda$ with $q_p \not= 0$. Then for $(x^1,x^2,x^3) = (t,x,y)$ we have
$q_{x^i}+q_p p_{x^i} = 0$, so $p_{x^i} = -q_{x^i}/q_p$. Substituting these into (\ref{M_1_covering})
yields~(\ref{M_2_covering}).

Our main observation in this paper is that the covering (\ref{M_2_covering}) allows the
generalization
\begin{gather}
q_t  = (u_x-q+v) q_x+v_x q,
\qquad
q_y  =  u_y q^{-1} q_x+v_y.
\label{generalized_M_2_covering}
\end{gather}
This system is compatible whenever the two-component system
\begin{gather}
u_{ty}  = (u_x+v) u_{xy}-u_y u_{xx},
\label{rdDym_2D_eq_1}
\\
v_{ty}  = (u_x+v) v_{xy}-u_y v_{xx} + v_x v_y
\label{rdDym_2D_eq_2}
\end{gather}
holds.  In other words, (\ref{generalized_M_2_covering}) is a covering for system
(\ref{rdDym_2D_eq_1}), (\ref{rdDym_2D_eq_2}).

\section{Reductions}
By the construction, we have the following reduction for system
(\ref{generalized_M_2_covering}):

{\it Reduction A}. Substituting for $v=0$ in equations  (\ref{rdDym_2D_eq_1}),
(\ref{generalized_M_2_covering}) gives equations (\ref{special_rdDym_eq}) and
(\ref{M_2_covering}), while (\ref{rdDym_2D_eq_2}) becomes an identity.

Also, we have three other reductions.

{\it Reduction B}.  If we put $v= -(\kappa^{-1}+1) u_x$, then
(\ref{rdDym_2D_eq_1}) gets the form
\begin{gather}
u_{ty} = -\kappa^{-1} u_x u_{xy} - u_y u_{xx},
\label{transformed_general_rdDym_eq}
\end{gather}
while (\ref{rdDym_2D_eq_2})  is its dif\/ferential consequence. The transformation
$u \mapsto - \kappa  u$ maps (\ref{transformed_general_rdDym_eq}) to (\ref{general_rdDym_eq}).
The corresponding reduction of (\ref{generalized_M_2_covering}) produces the covering of (\ref{general_rdDym_eq})
studied in \cite{Morozov2009, Pavlov2006}.

{\it Reduction C}. Taking $v=-u_x$ in (\ref{rdDym_2D_eq_1}), (\ref{rdDym_2D_eq_2}), we obtain
\[
u_{ty} = -u_y u_{xx}
\]
and its dif\/ferential consequence. Then we divide this equation by $u_y$, dif\/ferentiate w.r.t.~$y$ and put
$u_y = - e^w$. This gives the Boyer--Finley equation \cite{BoyerFinley}
\begin{gather}
w_{ty} = (e^w)_{xx}
\label{Boyer_Finley_equation}
\end{gather}
This equation is equation (\ref{Boyer_Finley_eq}) in a dif\/ferent notation.
Substituting for $q=e^p$ in the corresponding reduction of (\ref{generalized_M_2_covering}), we have the covering
\cite{KashaevSavelievSavelievaVershik,MalykhNutkuSheftel, Zakharov} for equation (\ref{Boyer_Finley_equation}):
\begin{gather*}
p_t  = w_t - e^p p_x,
\qquad
p_y  =  e^{w-p} (w_x-p_x).
\end{gather*}

{\it Reduction D.}
Finally, when we put $v = u_y-u_x$ into
(\ref{rdDym_2D_eq_1}) and (\ref{rdDym_2D_eq_2}), we get the equation
\[
u_{ty} =u_y \left(u_{xy}-u_{xx}\right)
\]
and its dif\/ferential consequence. Then for $u_y=e^w$ we have the deformed Boyer--Finley equation~\cite{Dryuma2007}
\begin{gather}
w_{ty} = \left(e^w\right)_{xy} - \left(e^w\right)_{xx},
\label{deformed_Boyer_Finley_eq}
\end{gather}
and the corresponding reduction of equations (\ref{generalized_M_2_covering}) with $q=e^s$ gives the covering
\begin{gather*}
s_t  = (e^s-e^w) s_x- w_t,
\qquad
s_y  =  e^w (s_x-w_x+w_y).
\end{gather*}
for (\ref{deformed_Boyer_Finley_eq}). This covering in other notations was found in \cite{Dryuma2007,Morozov2009}.

\section{B\"acklund transformations}

The substitution
\begin{gather}
u_x  = {-v+\frac{s_t}{s_x}},
\qquad
u_y  = {-\frac{e^{-r}}{s_x}},
\qquad
v_x  = {\frac{r_x  s_t}{s_x}-r_t},
\qquad
v_y  = {-\frac{e^{-r} r_x}{s_x}}
\label{two_component_Backlund_transformation}
\end{gather}
maps system (\ref{generalized_M_2_covering})  to
system
\begin{gather}
q_t  = {\left(\frac{s_t}{s_x}-q\right)  q_x+\left(\frac{s_t  r_x}{s_x}-r_t\right) q},
\qquad
q_y  = {-\frac{e^{-r}}{q  s_x} (q_x+r_x  q)}
\label{Bogdanov_covering}
\end{gather}
found in \cite{Bogdanov1}.  This system is the two-component generalization of the
covering
\begin{gather*}
q_t  = {\left(\frac{s_t}{s_x}-q\right)  q_x},
\qquad
q_y  = {-\frac{q_x}{q  s_x}}.
\end{gather*}
of equation  (\ref{universal_hierarchy_eq}).
The compatibility conditions for (\ref{Bogdanov_covering}) coincide with
(\ref{Bogdanov_system}).
Solving (\ref{two_component_Backlund_transformation}) for $s_t$, $s_x$, $r_t$, $r_x$ yields
\begin{gather}
s_t  = {-(u_x+v) \frac{e^{-r}}{u_y}},
\qquad
s_x  = {-\frac{e^{-r}}{u_y}},
\qquad
r_t  = {\frac{v_y}{u_y}},
\qquad
r_x  = {\frac{(u_x+v) v_y}{u_y}-v_x}.
\label{inverse_two_component_Backlund_transformation}
\end{gather}
This system is compatible whenever equations (\ref{rdDym_2D_eq_1}), (\ref{rdDym_2D_eq_2}) are satisf\/ied. Thus
equations (\ref{two_component_Backlund_transformation}) def\/ine a B\"acklund transformation from
(\ref{rdDym_2D_eq_1}), (\ref{rdDym_2D_eq_2}) to (\ref{Bogdanov_system}) with the inverse transformation
(\ref{inverse_two_component_Backlund_transformation}). In particular, when $v=0$ and $r=0$, we have a
B\"acklund transformation
\begin{gather*}
u_x  = {\frac{s_t}{s_x}},
\qquad
u_y  = {-\frac{1}{s_x}},
\end{gather*}
between (\ref{Pavlov_eq}) and (\ref{universal_hierarchy_eq}) with the inverse transformation
\begin{gather*}
s_t  = {-\frac{u_x}{u_y}},
\qquad
s_x  = {-\frac{1}{u_y}}.
\end{gather*}

\subsection*{Acknowledgments}
 I am very grateful to  M.V.~Pavlov and A.G.~Sergyeyev for the valuable  discussions.
Also I'd like to thank  M.~Marvan and A.G.~Sergyeyev for the warm hospitality in
 Mathematical Institute, Silezian University at Opava, Czech Republic, where this work was initiated and partially
 supported by the ESF project CZ.1.07/2.3.00/20.0002.

\pdfbookmark[1]{References}{ref}
\LastPageEnding

\end{document}